\input{aipcheck}
\documentclass[final]{aipproc}
\layoutstyle{6x9}

\usepackage{amsmath}

\begin{document}

\title{ 
Microscopic calculation of the decay of Jaffe-Wilczek tetraquarks,
and the Z(4433)
}

\author{M. Cardoso} {
address={Dep. F\'{\i}sica and Centro de F\'{\i}sica Te\'{o}rica de Part\'{\i}culas, Instituto Superior T\'{e}cnico,
Lisboa, Portugal}
}

\author{P. Bicudo}{
address={Dep. F\'{\i}sica and Centro de F\'{\i}sica Te\'{o}rica de Part\'{\i}culas, Instituto Superior T\'{e}cnico,
Lisboa, Portugal}
}


\begin{abstract}
Here the tetraquarks are studied a la Jaffe and Wilczek.
The decay width is fully computed with a microscopic quark
model, using the resonating group method.
\end{abstract}

\keywords{ Tetraquark, Confinement, Decay Width, Flip-flop  }
\classification{12.39.Mk, 12.39.Pn}

\maketitle

\section{Introduction}
\label{intro}
Recently the discovery of the charged ressonance state $Z(4433)$ by the Belle Collaboration
\cite{Belle} has led to the proposal that this state is indeed a radial excitation
of a tetratarquark of the form $uc\bar{d}\bar{c}$ \cite{Maiani}, being the particles
$X(3872)$ and $X(3876)$, discovered by Belle and Babar, the fundamental tetraquark states
( $uc\bar{u}\bar{c}$ and $dc\bar{d}\bar{c}$ ).

Others interpret this state as a scattering ressonance of the $D^*(2010) − \bar{D}(2420)$ system \cite{meng,rosner}
while still others interpret this as a threshold cusp in the same system \cite{bugg}.

The $Z(4433)$ was detected by the decay $Z(4433) \rightarrow \pi^{\pm} \psi$, but if the model of the tetraquark is
correct, the decay $Z(4433) \rightarrow D^{\pm} D^0$ could also be possible.

\section{Spectrum and wavefunctions}

We start to study the tetraquark system, by using the flux tube model, in wich we have a potential given by

\begin{equation}
	V(r_1,r_2,r_3,r_4) = V_{Coulomb} + V_{Conf}
\end{equation}

The term $V_{Coulomb}$ is the Coulomb part of the interaction corresponding to one gluon exchange,
and $V_{Conf}$ is the confining part of the potential.

The Coulomb interaction should be proportional to $t_1^a t_2^a$, where $t_1$ and $t_2$ are the Casimir operators
of the two particles.
Since the two quarks are in an antitriplet state we get $(t_1+t_2)^2 = t_1^2 = t_2^2 = \frac{4}{3}$, wich gives
$t_1 \cdot t_2 = - 2 / 3$ so the quark-quark potential is
\begin{equation}
	V_{qq}(r) = - \frac{2}{3} \frac{\alpha_s}{r} 
\end{equation}
For the quark-antiquark Coulomb potential, since a quark and two antiquarks form an antitriplet, we get a
factor $-1/3$ for the quark-antiquark interaction
\begin{equation}
	V_{q\bar{q}}(r) = - \frac{1}{3} \frac{\alpha_s}{r} 
\end{equation}

As for the confining part of potential it is given by
\begin{equation}
	V_{conf}( r_1, r_2, r_3, r_4 ) = \sigma L_{min}( r_1, r_2, r_3, r_4 )
\end{equation}

\begin{figure}
\centering
\includegraphics[width=0.35\textwidth]{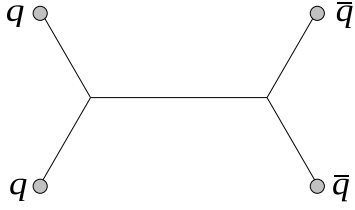}
\caption{ String picture for the tetraquark. }
\label{tetraquark}
\end{figure}

Where $L_{min}$ is the minimum distance that links the four particles as shown in figure \ref{tetraquark}

The potential is given by
\begin{equation}
	V(\mathbf{r}_1,\mathbf{r}_2,\mathbf{r}_3,\mathbf{r}_4) = C
	- \frac{4}{3} \alpha_s ( \frac{1}{2} \frac{1}{r_{12}} + \frac{1}{2} \frac{1}{r_{34}}
	+ \frac{1}{4} \frac{1}{r_{13}} + \frac{1}{4} \frac{1}{r_{14}} + \frac{1}{4} \frac{1}{r_{24}} )
	+ \sigma L_{min} \big( \mathbf{r}_1, \mathbf{r}_2, \mathbf{r}_3, \mathbf{r}_4 \big)
\end{equation}

We should solve the schroedinger equation for this potential, but because the form of the confining part of the potential is cumbersome, we approximate our system by a system of a diquark and a di-antiquark, so that we only have the interaction between quarks ( antiquarks ) in the diquark ( diantiquark ) and the interaction diquark-diantiquark. Our simplified potential is given by

\begin{eqnarray}
	V ( \mathbf{r}_{u}, \mathbf{r}_{c}, \mathbf{r}_{\bar{c}}, \mathbf{r}_{\bar{d}} ) &=&
	V_{diq}( \mathbf{r}_{u}, \mathbf{r}_{c} ) + V_{diq}( \mathbf{r}_{\bar{c}}, \mathbf{r}_{\bar{d}} )
	+ V_{diq-diq}( \mathbf{R}_{uc}, \mathbf{R}_{\bar{c}\bar{d}} ) \\ \nonumber
	&=& \sigma | \mathbf{r}_{u} - \mathbf{r}_{c} | + \sigma | \mathbf{r}_{\bar{d}} - \mathbf{r}_{\bar{c}} |
	+ \sigma | \mathbf{R}_{uc} - \mathbf{R}_{\bar{c}\bar{d}} |
\end{eqnarray}
where $\mathbf{R}_{uc}$ and $\mathbf{R}_{\bar{c}\bar{d}}$ are the centers of mass of the $uc$ and $\bar{d}\bar{c}$ system.
\begin{equation}
	V_{diq}( \mathbf{r}_{1}, \mathbf{r}_{2} ) = C_{diq} - \frac{2}{3} \frac{\alpha_s}{ | \mathbf{r}_1 - \mathbf{r}_2 | } 
	+ \sigma | \mathbf{r}_1 - \mathbf{r}_2 |
\end{equation}
\begin{equation}
	V_{diq-diq}( \mathbf{r}_{1}, \mathbf{r}_{2} ) =
	C_{diq-diq} - \frac{4}{3} \frac{\alpha_s}{ | \mathbf{r}_1 - \mathbf{r}_2 | } + \sigma | \mathbf{r}_1 - \mathbf{r}_2 |
\end{equation}

So we define the coordinates
\begin{eqnarray}
	\mathbf{R} =& \frac{ m_u \mathbf{r}_{u} + m_c \mathbf{r}_{c} + m_u \mathbf{r}_{\bar{d}} + m_c \mathbf{r}_{\bar{c}} }
		{ 2 m_u + 2 m_c } \\
	\boldsymbol{\rho}_{uc} =& \mathbf{r}_{u} - \mathbf{r}_{c} \\
	\boldsymbol{\rho}_{\bar{d}\bar{c}} =& \mathbf{r}_{\bar{d}} - \mathbf{r}_{\bar{c}} \\
	\mathbf{r}_{T} =& \mathbf{R}_{uc} - \mathbf{R}_{\bar{d}\bar{c}}
\end{eqnarray}

Then we could also write the kinetical energy operator as

\begin{equation}
	\hat{T} = \frac{\mathbf{P}^2}{2M}
		+ \frac{\mathbf{p}_1^2}{2 \mu_1} + \frac{\mathbf{p}_2^2}{2 \mu_2}
		+ \frac{\mathbf{p}^2}{2\mu}
\end{equation}

Writing the wavefunction as
\begin{equation}
	\Psi( \mathbf{r}_{u}, \mathbf{r}_{c}, \mathbf{r}_{\bar{c}}, \mathbf{r}_{\bar{d}} ) =
	\psi_{CM}( \mathbf{R} )
	\psi_{uc}( \boldsymbol{\rho}_{uc} )
	\psi_{\bar{d}\bar{c}}( \boldsymbol{\rho}_{\bar{d}\bar{c}} )
	\phi_{T}( \mathbf{r}_{T} )
\end{equation}

We arrive at three independent Schrodinger equations
\begin{equation}
	- \frac{\hbar^2}{2m} \nabla^2 \psi + V \psi = \epsilon \psi
\end{equation}
In this work we only care about s wave particles, so we expand the wavefunctions in the s-wave solutions of the
harmonic oscilator
\begin{equation}
	\psi_i ( \boldsymbol{\rho}_i ) = \sum_{n} A_{n} \frac{ H_{2 n +1} ( \rho_i ) }{ \rho_i } e^{ - \frac{\rho^2}{2 a} }
	\label{expansion}
\end{equation}
We do the same expansion to calculate the meson wavefunctions, wich we use latter to calculate the decay width.
We use the parameters $ \sigma = (440 MeV)^2$, $ \frac{4}{3} \alpha_s = 0.27$ and fix the constant so that we get a mass of $4433 MeV$ for the first radial excitation of the tetraquark, and we get the value $C = -2328 MeV$.
This is a standard procedure for mesons, we choose the potential constant to get the correct masses. The results are given in table \ref{mesonlev}.

\begin{table}[!ht]
	\begin{tabular}{ccc}
		\hline
		Particle & $C(MeV)$ & $M(MeV)$ \\
		\hline
		$\pi$ & -1557 & 140 \\
		$\eta_c$ & -680 & 2980 \\
		$J/\Psi$ & -797 & 3097 \\
		$D$ & -976 & 1867 \\
		$Z(1S)$ & -2328 & 3913 \\
		$Z(2S)$ & -2328 & 4433 \\
		$Z(3S)$ & -2328 & 4830 \\
		\hline
	\caption{ Masses and constant $C$ for different particles }
	\end{tabular}
	\centering
	\label{mesonlev}
\end{table}

\section{Decay with microscopic quark exchange}

Now we study the decay of the tetraquark system into a two meson system. Note that the potential is different in the tetraquark and in the meson-meson channels. We assume the "true" potential of the system is the the flip-flop potential
\begin{equation}
	V_{FF}( \mathbf{r}_{u}, \mathbf{r}_{c}, \mathbf{r}_{\bar{d}}, \mathbf{r}_{\bar{c}} ) =
	\min( V_{T}, V_{\pi \psi}, V_{DD} ) =
	\min( V_{T}, V_{\pi} + V_{\psi}, V_{D^0} + V_{D^{+}} )
\end{equation}
Where $V_T$ is the tetraquark potential, given above, $V_{\pi}$, $V_{\psi}$, $V_{D^{0}}$ and $V_{D^{+}}$ are the quark-antiquark potentials in the respective mesons, wich is of the funnel type
\begin{equation}
	V_i( \mathbf{r}_i ) = C_i - \frac{4}{3} \frac{\alpha_s}{r_i} + \sigma r_i
\end{equation}
The difference $ V_{FF} - V_{T} $ is the perturbation that causes the decay of the tetraquark into two mesons. The full hamiltonean is given by
\begin{equation}
	\hat{H} = \sum_i T_i(\mathbf{p}_i) +
	V_{FF}( \mathbf{r}_{u}, \mathbf{r}_{c}, \mathbf{r}_{\bar{d}}, \mathbf{r}_{\bar{c}} )
\end{equation}
We could rewrite the hamiltonean as $H = H_1 + V_1$ with $V_1 = V_{FF} - V_T$ and $V_2 = V_{FF} - V_{\pi\psi}$.
Now we calculate the matrix elements
\begin{equation}
	\langle T | H | T \rangle = M_T^0 + \langle T | V_1 | T \rangle
\end{equation}
Note that we need to redefine the potential so that $M_T = M_T^0 + \langle V_1 \rangle$.
We have, also
\begin{eqnarray}
	\langle \pi \Psi | H | T \rangle &=&
	M_T^0 \langle \pi \Psi | T \rangle + \langle \pi \Psi | V_1 | T \rangle \\
	\langle \pi \Psi | H | \pi \Psi \rangle &=&
	M_{\pi} + M_{\Psi} + \langle \pi \Psi | V_2 | \pi \Psi \rangle
\end{eqnarray}
To compute the decay of the tetraquark in two mesons we have to solve the equation
\begin{equation}
	\left( \begin{array}{cc}
	M_T^0 + \langle V_1 \rangle_T & M_T^0 e^* \langle \phi | + \nu^* \langle \chi | \\
	M_T^0 e | \phi \rangle + \nu | \chi \rangle & M_{\pi} + M_{\psi} + T + \langle V_2 \rangle_{\pi\psi}
	\end{array} \right)
	\left( \begin{array}{c} 1 \\
	| \pi-\psi \rangle
	\end{array} \right) =
	E S
	\left( \begin{array}{c} 1 \\
	| \pi-\psi \rangle
	\end{array} \right)
\end{equation}
where
\begin{equation}
	S = \left( \begin{array}{cc}
	1 & e^* \langle \phi | \\
	e | \phi \rangle & 1
	\end{array} \right)
\end{equation}
\begin{equation}
	\nu \langle \mathbf{r}_{\pi\psi} | \chi \rangle =
	\frac{1}{8} \int d^3 \boldsymbol{\rho}_{\pi} d^3 \boldsymbol{\rho}_{\psi}
	\psi_{uc}( \boldsymbol{\rho}_{uc} )^* \psi_{\bar{d}\bar{c}}( \boldsymbol{\rho}_{\bar{d}\bar{c}} )^*
	\phi_{T}( \mathbf{r}_{t} ) V( \boldsymbol{\rho}_{\pi}, \boldsymbol{\rho}_{\psi}, \mathbf{r}_{\pi\psi} )
	\psi_{\pi}( \boldsymbol{\rho}_{\pi} ) \psi_{\psi}( \boldsymbol{\rho}_{\psi} )
\end{equation}
\begin{equation}
	e \langle \mathbf{r}_{\pi\psi} | \phi \rangle =
	\frac{1}{8} \int d^3 \boldsymbol{\rho}_{\pi} d^3 \boldsymbol{\rho}_{\psi}
	\psi_{uc}( \boldsymbol{\rho}_{uc} )^* \psi_{\bar{d}\bar{c}}( \boldsymbol{\rho}_{\bar{d}\bar{c}} )^*
	\phi_{T}( \mathbf{r}_{t} )
	\psi_{\pi}( \boldsymbol{\rho}_{\pi} ) \psi_{\psi}( \boldsymbol{\rho}_{\psi} )
\end{equation}

The term $e | \phi \rangle$ comes from the nonorthogonality between the tetraquark state and the
meson-meson state. If the states where orthogonal the $S$ matrix would be the identity.

We have the equations
\begin{equation}
	M_T - E + ( e^* M_T^0 - e^* E ) \langle \phi | \pi - \psi \rangle + \nu^* \langle \chi | \pi - \psi \rangle = 0
\end{equation}
\begin{equation}
	( M_T^0 - E ) e | \phi \rangle + \nu | \chi \rangle + ( M_{\pi} + M_{\psi} + T + \langle V_2 \rangle_{\pi\psi} - E )
	| \pi - \psi \rangle = 0
\end{equation}

Now eliminating $| \pi - \psi \rangle$ we get
\begin{equation}
	E - M_{T} =
	- \int \frac{d^3 \mathbf{p}}{(2\pi)^3} \frac{ \chi^*(\mathbf{p}) \chi(\mathbf{p}) }
	{ M_{\pi} + M_{\psi} + T(\mathbf{p}) + \langle V_2 \rangle_{\pi\psi} - E - i \epsilon }
\end{equation}
where
\begin{equation}
	\chi(\mathbf{p}) = ( M_T^0 - E ) e \langle \mathbf{p}_{\pi\psi} | \phi \rangle +
	\nu \langle \mathbf{p}_{\pi\psi} | \chi \rangle
\end{equation}

To calculate the decay width, we make the replacement $E \rightarrow E - i \frac{\Gamma}{2}$
in the left side of the equation and make $E \simeq M_T$ in the right side. Also we neglect the term
$\langle V_2 \rangle_{\pi\psi}$ wich corresponds to a short range iteraction between the mesons.

For the terms $ \nu \langle \mathbf{p}_{\pi\psi} | \chi \rangle  $ and
$ e \langle \mathbf{p}_{\pi\psi} | \phi \rangle $ we use the eq. $\ref{expansion}$ for all the wavefunctions,
so that $ \nu \langle \mathbf{p}_{\pi\psi} | \chi \rangle  $ is given by
\begin{eqnarray}
	\nu \langle \mathbf{p}_{\pi\psi} | \chi \rangle &=&
	N \int d^3 \mathbf{r}_{\pi\psi} d^3 \boldsymbol{\rho}_{\pi} d^3 \boldsymbol{\rho}_{\psi}
	e^{ - \frac{1}{2} C_{\pi\pi} \rho_{\pi}^2 } e^{ - \frac{1}{2} C_{\psi\psi} \rho_{\psi}^2 }
	e^{ - \frac{1}{2} C_{r} \rho_{\pi\psi}^2 }
	e^{ -C_{\pi\psi} \boldsymbol{\rho}_{\pi} \cdot \boldsymbol{\rho}_{\psi} - i \mathbf{p}_{\pi\psi} \cdot \mathbf{r}_{\pi\psi} } \\ \nonumber &&
	P_1( \boldsymbol{\rho}_{uc} ) P_2( \boldsymbol{\rho}_{\bar{c}\bar{d}} ) P_3( \boldsymbol{\rho}_{\pi} )
	P_4( \boldsymbol{\rho}_{\psi} ) P_5( \mathbf{r}_T )
	V( \boldsymbol{\rho}_{\pi}, \boldsymbol{\rho}_{\psi}, \mathbf{r}_{T} )
\end{eqnarray}
where the $P_i$ are polynomials. $\boldsymbol{\rho}_{uc}$, $\boldsymbol{\rho}_{\bar{d}\bar{c}}$ and $\mathbf{r}_T$
are given in terms of $\boldsymbol{\rho}_{\pi}$, $\boldsymbol{\rho}_{\psi}$ and $\mathbf{r}_{\pi\psi}$, by
\begin{eqnarray}
	\boldsymbol{\rho}_{uc} &=& \frac{1}{2}( \boldsymbol{\rho}_{\pi} - \boldsymbol{\rho}_{\psi} ) + \mathbf{r}_{\pi\psi} \\
	\boldsymbol{\rho}_{\bar{d}\bar{c}} &=& \frac{1}{2}( \boldsymbol{\rho}_{\psi} - \boldsymbol{\rho}_{\pi} ) + 
	\mathbf{r}_{\pi\psi} \\
	\mathbf{r}_{T} &=& \mathbf{R}_{uc} - \mathbf{R}_{\bar{d}\bar{c}}
	= \frac{ m_u }{ m_u + m_d } \boldsymbol{\rho}_{\pi} + \frac{m_c}{m_u + m_c} \boldsymbol{\rho}_{\psi}
\end{eqnarray}

For $ e \langle \mathbf{p}_{\pi\psi} | \phi \rangle$, we just replace the potential by $1$.
Since the weight of the integral is gaussian, we use the Monte Carlo method to evaluate this nine-dimensional integral.
So
we calculate the width for the decay of the $Z(4433)$ ( assuming it is a tetraquark in the $2S$ state ) in to $\pi \psi(1S)$
and $\pi \psi(2S)$, and obtain ( preliminary results ) $\Gamma( \pi \psi(1S) ) = 0.2 MeV$ and
$\Gamma( \psi(2S) ) = 4.6 MeV$.

\section{Conclusions}

The result is somewhat smaller than the experimental result for the decay of the $Z(4433)$ wich has a width of
$\Gamma = 44^{+17+13}_{-13-11} MeV$ \cite{rosner}, but this result is only preliminary. Also, we don't include the
spin effects, and it is known that the hyperfine splitting effect could increase the decay width.

Even though, we think that our method is quite powerfull and general, and we expect to get more accurate and
conclusive results using this method.


\end{document}